\documentclass{article}
  \usepackage{graphicx}

\def\beq{\begin{equation}}
\def\eeq{\end{equation}}
\def\pdr{\partial}
\def\pd{\partial}

\def\m#1{$#1$}
\def\half{{1\over 2}}
\usepackage{amssymb}
\def\beqs{\begin{eqnarray}}
\def\eeqs{\end{eqnarray}}
\def\[{\left[}
\def\]{\right]}
\def\({\left(}
\def\){\right)}

\begin{document}

\centerline{\bf Quantum Thermodynamics of Non-Ideal Gases}

\begin{center}{\sf A. Coutant\footnote{ENS de Lyon, 
        46, All\'ee d'Italie, 69364 Lyon CEDEX 07, France;  antonin.coutant@ens-lyon.fr}
 and S. G. Rajeev\footnote{rajeev@pas.rochester.edu}}\\
Department of Physics and Astronomy\\ 
Department of Mathematics\\
University of Rochester\\
  Rochester NY 14627
\end{center}


\centerline{\bf Abstract}
We show that when  the thermal wavelength is comparable to the spatial size of a system, thermodynamic observables  like Pressure and Volume have quantum fluctuations that cannot be ignored. They are now represented by operators; conventional (classical)  thermodynamics is no longer applicable.  We continue the work in earlier papers where quantization rules for thermodynamics were developed by analogy with optics and mechanics, by working out explicitly the quantum theory of van der Waals gases. We find a wave equation satisfied by the thermodynamic wave function as well as  solutions  ({\em coherent states}) that are centered at  the classical equations of state. The probability of departure from the classical theory is dependent on a  parameter $\sigma$  which is a property of the gas molecule.
  \pagebreak
    
\section{Introduction}

Usually we use quantum mechanics to describe small systems like individual atoms and thermodynamics to describe large systems like gases. But there are a few examples where quantum mechanics is needed for systems that are large enough to be also thermodynamic:  a well-established example would be a superconductor.  We also now have experiments with quantum gases at temperatures small enough that the thermal wavelength is of the order the distance between atoms, leading to Bose condensation  \cite{bosecond}. Another leap in which the thermal wavelengths are comparable to the size of the trap containing the gas can be contemplated. That will bring in a new set of quantum effects.

In this paper we will continue  to develop a quantum theory of thermodynamics, begun in Ref.  \cite{QTh}.  By this we do not mean just quantum statistical mechanics. For example, in deriving the partition function of the Bose gas, we use quantum mechanics. However,  the thermodynamic quantities such as Pressure or Volume are then treated as numbers not operators. By quantum thermodynamics we mean\footnote{Another theory, also called sometimes  quantum thermodynamics has already been proposed in series of interesting papers \cite{Nieuwenhuizen}. Our work does not seem to be related to it .} a situation where thermodynamic  variables  have quantum fluctuations and therefore  are to be treated as operators.  Recall that in classical thermodynamics, as in classical mechanics, observables come in conjugate pairs such as $P,V$ or $T,S$.

In the  earlier paper  \cite{QTh} we have suggested a thermodynamic uncertainty principle for quantum gases, bounding the product of quantum fluctuations in thermodynamically conjugate variables such as $P$ and $V$. If the  gas is cold enough,  and the trap  is small enough,  that the thermal wavelength of the particles is comparable to the diameter of the trap, we can no longer treat $P$ and $V$  as classical (commuting) observables. They are operators acting on a thermodynamic analogue of the quantum mechanical wavefunction. The equation of state of the thermodynamics system is not any more a relation between quantities such as $P,V,T$: such operator relations would be inconsistent. Instead we have a differential equation satisfied by a thermodynamic analogue of the wave function.

What is this differential equation, the thermodynamic analogue of the \linebreak Schr\"odinger equation? The Schrodinger equation of quantum mechanics reduces to the Hamilton-Jacobi equation of classical mechanics. Thus a way to discover the quantum thermodynamic wave equation is to first find the Hamilton-Jacobi formulation of thermodynamics. This was accomplished in Ref.  \cite{H-J}. 

In this paper we will study the resulting quantum wave equation further using as an example the van der Waals theory of  non-ideal gases. We chose this example because it is known to be a good model of real gases: the  equations of state of hundreds of gases  have been fit to van der Waals (vdW)  theory, with  widely available tables  \cite{vdWTable} of the parameters for each molecule. Moreover, it turns out that the vdW is a kind of `integrable system': by a change of variable we discovered in  \cite{H-J} its canonical equations can be solved analytically in closed form. Thus this system provides the correct balance between mathematical accessibility and physical realism.

We will work within the canonical rather than the micro-canonical ensemble. This means that our approach will not be valid near a phase transition to a condensate. A deeper analysis including chemical potential as a thermodynamic variable is conceptually similar, but will be more complicated because of the extra degree of freedom. We hope to study that extension in a later publication.

\section{The Thermodynamic Uncertainty Principle }

Let us review the argument for an uncertainty principle for thermodynamically conjugate variables. Consider a gas of molecules contained by some potential well acting as a `container' or `trap'. Classically there is a well defined meaning to the position of the `wall' of the container: it is the point at which the particle is reflected back into the trap, or the location of the classical turning point. But, in quantum mechanics things are not so clear-cut: the particle has  a small probability to tunnel past the turning point, or to be reflected before the turning point is reached.  Thus the linear size of the container $L$ has an uncertainty $\Delta L$ due to quantum fluctuations, with a resulting uncertainty of order $\Delta V=A\Delta L$ in the volume. 

There is also an uncertainty in the Pressure. Pressure is the force exerted per unit volume on the wall when particles and reflected back.
This is of order
\beq
P=2p \rho v .
\eeq
Here $p$ is the component of momentum normal to the wall. Also,  $\rho v $ is the number of collisions (per unit time per unit area) with the wall: the product of number density and average velocity.

 In quantum mechanics, the momentum of the particle  and the position  have quantum fluctuations, with the uncertainties bounded by 
 \beq
 \Delta p\Delta L \geq {\hbar\over 2}.
 \eeq
This translates to a quantum uncertainty in pressure and $L$:
\beq
 \Delta P\Delta L \geq \hbar \rho v.
\eeq

 We should verify  that for most conventional applications of thermodynamics, this quantum uncertainty can be neglected. To get an order of magnitude of this error, we can use the ideal gas equations of  state
 \beq
 PV=Nk_BT,
  \eeq
where $k_B$ is Boltzmann's constant and $N$ the total number of particles. Or,
\beq
{\rho\over P}={1\over k_B T}.
\eeq
Thus
\beq
 {\Delta P\over P}\Delta L \geq \hbar  {v\over k_B
T}.
\eeq
On the other hand, by  equipartition of energy, ${1\over 2} k_B
T$ is the contribution to kinetic energy due to motion normal to the wall; this is equal to the momentum  times the velocity normal to the wall. So
\beq
 {\Delta P\over P}\Delta L \geq {\hbar\over p_T}, 
\eeq
where
\beq
p_T=\sqrt{mk_BT}
\eeq
We have the equivalent forms 
\beq
 {\Delta P\over P}{\Delta L \over L} \geq {\sqrt{2\pi} \Lambda\over L}
\eeq
and since $V\propto  L^3$, 
\beq
{\Delta P\over P}{\Delta V \over V} \geq  {3\sqrt{2\pi}}{ \Lambda\over L}
\eeq
where $\Lambda={h\over \sqrt{2\pi mk_BT}}$ is the thermal  wavelength of a particle in the gas \cite{pathria}. 

This is the thermodynamic uncertainty principle. The quantum bose gases in magneto-optical traps have achieved very  low temperatures already \cite{bosecond}; it might be possible to observe such fluctuations experimentally in a later generation of such experiments.

Thus, if the size  of the trap is comparable to the thermal  wavelength, 
\beq
{3\sqrt{2\pi}}\Lambda\geq L.   \label{quantcond}
\eeq
pressure and volume cannot be treated as commuting observables. They are quantum operators acting on a thermodynamic analogue of the wave-function of quantum mechanics.

The equation of state therefore cannot be a relation among $P,V,T$: such an operator relation would be inconsistent. Instead we must have a wave equation satisfied by the thermodynamic wave function.

The above condition is to be compared with the much weaker condition at which the quantum statistical effects (e.g., Bose statistics) become significant:
\beq
\Lambda\geq \left(V\over N\right)^{1\over 3}
\eeq
That is, the thermal wavelength is greater than the inter-particle distance. 

It is unlikely that a weakly interacting Fermionic system will satisfy the condition (\ref{quantcond}): the wavelength at the Fermi surface is likely to be too small. However, composites of fermions may have this behavior. In any case, weakly non-ideal  bosonic gases would seem to be the best candidates. 

\section{Equation of State of a Non-Ideal Gas}

For an ideal gas
\beq
P={T\over v}
\eeq
where \m{v} is   volume per particle
\beq
v={V\over k_B N}
\eeq
and \m{k_B} is the Boltzmann constant\footnote{ It will be convenient to scale extensive quantities
\beq
s={S\over k_B N},\quad u={U\over k_B N}
\eeq
etc.
 to simplify formulas. Conventional units can be restored at the end using simple dimensional analysis.}.

The standard  classical model for non-ideal gases is the van der Waals theory. It describes very well the behavior of many materials in the real world. Also, it gives  a good description of the liquid-gas transition away from the critical point.

In  the van der Waals theory  \cite{huang,pathria} a  molecule is assumed to have  a finite volume $b$, so that the actual volume available for the gas is $N[v-b]$. Moreover, there is a force between pairs of molecules. In the  bulk the net force is zero as there are the same number of molecules in all directions. But at the wall, there is a net force normal to the wall (there being no gas molecules on the other side) so the pressure  is changed  by an amount proportional to the square of density. If the force between the molecules is attractive this is a decrease in pressure; otherwise it is an increase. Thus we are led to a modified equation of state
\beq
P={T\over v-b}-{a\over v^2};
\eeq
the last term being proportional to the density of {\em pairs} of particles, or to the square of number density.

Another equation of state is the formula for $u={U\over k_BN}$, the internal energy per particle
\beq
u={3\over 2}T-{a\over v}.
\eeq
The last term is the extra energy each particle gains through interactions with its neighbors. 
Thus, we can regard the departure from ideal gas behavior as the addition of constants \m{-a,b} to the variables \m{v, uv}. A thermodynamic description in which these are used as the co-ordinates, will be especially convenient.

In the original van der Waals theory, the molecules attract at large distances so that $a>0$. But  both signs for $a$ are  interesting from a modern perspective.

These arguments apply also  to gases at low enough temperatures  to be quantum.  For example, in the Lee-Huang-Yang \cite{LHY}  theory of dilute non-ideal Bose  gases, the equation of state is of the above form with $b=0$ and  $a<0$: the negative sign being because the pairwise force is repulsive.  More precisely, 
\beq
a=-8\pi {\hbar^2\over 2mk_B}a_s,
\eeq
where $a_s$ is the scattering length\footnote{$k_B$ appears because our  definition   of energy includes it as a factor: $u={U\over Nk_B}$. We have to convert from the units used in Ref.  \cite{LHY} to ours. 
}. If the scattering length is positive the force is repulsive. More generally, $a$ is related to the energy of the lowest resonance in the two-body elastic scattering of two molecules.   We will thus use the van der Waals as our generic model of a non-ideal gas even in the quantum regime.

A gas has a thermodynamic phase space with five co-ordinates $(u,T,s,P,v)$. The first and second laws of thermodynamics require the relation
\beq
du=Tds-Pdv
\eeq
among the infinitesimal variations of these quantities. Here we see that thermodynamic variables appear in conjugate pairs such as $(T,s)$ and $(P,v)$. Because of the odd-dimensionality,  there appears to be an unpaired variable $u$. However, by rewriting the relation as
\beq
ds={du\over T}+{P\over T}dv
\eeq
for example, we can see that there is nothing special about $u$ as a thermodynamic co-ordinate. As explained elsewhere  \cite{H-J, Hermann,  Ruppeiner} classical thermodynamics has a natural formulation in terms of contact geometry, the odd dimensional analogue of the symplectic geometry that describes mechanics. The quantities $(u,v)$ are canonically conjugate to 
\beq
p_u={1\over T},\quad p_v={P\over T}.
\eeq

Among the five variables there are three relations, the equations of state of the gas. These relations are most simply described by giving the entropy  $s$ as a function of $(u,v)$: the {\em fundamental relation}. For the van der Waals gases, this function also depends on two additional parameters $(a,b)$ which describe the departure of the gas from being ideal:
\beq
s=s(u,v; a,b)
\eeq
Given this  function, the remaining variables are given by the derivatives
\beq
{1\over T}=\left({\pd s\over \pd u}\right)_v, \quad {P\over T} =\left({\pd s\over \pd v}\right)_u  .
\eeq
For fixed $(a,b)$, this determines a two dimensional surface in the five-dimensional thermodynamic phase space: the {\rm Lagrange sub-manifold} of this particular gas.

The fundamental relation  of  case of a van der Waals gas is, explicitly,
\beq
s=\log(v-b)-{3\over 2}\log v +{3\over 2}\log[uv+a]. \label{fundamental}
\eeq
A simple calculation  of the derivatives shows that we get the van der Waals formula for $P$ upon eliminating $u$.

\section{Hamilton-Jacobi Theory of van der Waals Gases}
 
 It is possible to `unify'  \cite{H-J} the van der Waals model of gases, so that the fundamental relation $s(u,v;a,b)$ is the solution of a common differential equation: the parameters $a,b$ appear as constants of integration. This is the analogue, in thermodynamics, of the Hamilton-Jacobi formulation of mechanics or optics. In the usual picture where $u,v$ are used as co-ordinates,  entropy would  the analogue of the eikonal. 

By eliminating $a,b$ we can get a relation among the five thermodynamical co-ordinates:
\beq
F(s,u,v, p_u,p_v)=0 \label{hypersurface}
\eeq
If we make the canonical substitutions
\beq
p_u={\pdr s\over \pdr u},\quad p_v={\pd s\over \pdr v}
\eeq
we get   the  thermodynamic  analogue of the Hamilton-Jacobi equation:
\beq
F(s,u,v,{\pd s\over \pd u},{\pd s\over \pd v})=0
\eeq

All solutions of such a first order PDE are determined by a  fundamental solution\footnote{See  \cite{CourantHilbert2} where it is called the `complete integral'. We propose the name `fundamental solution' instead, which fits better with the term `fundamental relation' as used in thermodynamics.}  that depends on two constants of integration. It   is in some ways analogous to the plane wave solution  of a wave equation; every other solutions can be expressed as a linear superposition of plane waves. All other solutions of a first order PDE are obtained from the fundamental solution by the steepest descent approximation to such a superposition: by an extremization.

The explicit form of the equation for the hypersurface of the  vdW famility of  gases (obtained by elininating $a,b$ from the equations of state)  is  \cite{H-J}:
\beq
\left[vp_v-up_u+{3\over 2}
\right]^2p_u^3={27\over 8}v^2e^{-2s}
\eeq
so that the H-J equation is 
\beq
\left[v\left({\pd s\over\pd v} \right)-u\left({\pd s\over \pd u}\right)+{3\over 2}
\right]^2\left[{\pdr s\over \pdr u}\right]^3={27\over 8}v^2e^{-2s}
\eeq

 We will now  solve the Hamilton-Jacobi equation and recover the fundamental relation (\ref{fundamental}). This exercise will also help us identify the natural variables of the system,  in which the quantization will be easier. 
\section{  Normal  Co-ordinates}
It will be convenient to change to some new variables (rather like normal co-ordinates of a mechanical system) $(\phi,q_1,q_2,p_1,p_2)$ which still satisfy  thermodynamic relation
 \beq
 d\phi=p_1dq_1+p_2dq_2.
 \eeq
 That is, a   Legendre Transformation, the thermodynamic analogue of a  canonical transformation of classical mechanics. We can choose the transformation  such that the H-J equation becomes independent of $\phi$. This will make it possible to find the fundamental solution  by separation of variables and recover the van der Waals relation. These variables were discovered in  \cite{H-J}, so we just give the answer and show how the reader can verify that it has the correct properties:

\beq
q_1=v,\quad q_2={3\over 2} uv,\quad
\eeq
and
\beq
 p_1=\left[vp_v-up_u+{3\over 2}\right]\left(v^{-1}e^s
\right)^{2\over 5} ,\quad  p_2={2\over 3 }p_u\left(v^{-1}e^s\right)^{2\over 5},
\eeq
\beq
 \phi={5\over 2}v^{3\over 5}e^{{2\over 5}s}
\eeq
It is straightforward (and not too  tedious) to check that this  is indeed a Legendre Transformation:
\beq
d \phi=p_1dq_1+p_2dq_2 \Leftrightarrow ds=p_udu+p_vdv.
\eeq

These variables have been chosen such that, the hypersurface equation (\ref{hypersurface})  is simply 
\beq
p_1^2p_2^3=1.
\eeq
The H-J equation is then obtained by the canonical replacement
\beq
p_1={\pdr \phi\over \pd q_1},\quad p_2={\pdr \phi\over \pd q_2}
\eeq
to get 
\beq
\left({\pd \phi\over \pd q_1}\right)^2\left({\pd \phi\over \pd q_2}\right)^3=1.
\eeq

We can now    solve this PDE by separation of variables, and recover the equations of state of the vdW gas:
\beq
\phi(q_1,q_2)=\phi_1(q_1)\phi(q_2)
\eeq
so that 
\beq
[\phi_1'(q_1)]^2[\phi_2(q_2)]^2[\phi_2'(q_2)]^3[\phi_1(q_2)]^3=1.
\eeq
Or,
\beq
\phi_1'=C_1\phi_1^{-{3\over 2}},\quad \phi_2'=C_2\phi_2^{-{2\over 3}}
\eeq
for some separation constants satisfying 
\beq
C_1^2C_2^3=1.
\eeq These ODE are easy enough to solve:
\beq
\phi_1(q_1)=K_1(q_1-q_1')^{2\over 5},\quad \phi_2(q_2)=K_2(q_2-q_2')^{3\over 5}
\eeq

Here, \m{q_1',q_2'} are constants of integration. The product  \m{K_1K_2} is  determined through
\beq
C_1={2\over 5}K_1^{5\over 2},\quad C_2={3\over 5}K_2^{5\over 3}.
\eeq
so that 
\beq
2^2 3^3 [K_1K_2]^5=5^5\Rightarrow 
K_1K_2=5\  2^{-{2\over 5}}\  3^{-{3\over 5}}
\eeq
Thus
\beq
\phi(q_1,q_2)= \alpha (q_1-q_1')^{2\over 5}(q_2-q_2')^{3\over 5}
\eeq
where
\beq
\alpha= {5\over (2^2 3^3)^{1\over 5}}.
\eeq

Recalling  the definitions  $ \phi={5\over 2} v^{3\over 5}e^{{2\over 5}s}, q_1=v, q_2={2\over 3}uv$   we get the familiar formula (\ref{fundamental})  for the entropy\footnote{Up to an additive  numerical constant. } of a vdW gas:
\beq
s=\log\left[(v-b)\left(u+{a\over v} \right)^{3\over2}
\right]
\eeq
with the identification of the constants of integration:
\beq
b=q_1',\quad a=-{2\over 3}q_2'.
\eeq
Thus our Hamilton-Jacobi equation `unifies' the vdW gases: the fundamental relation of different gases are solutions of the same PDE, with the parameters describing the gases appearing as constants of integration.

Using $p_1={\pdr \phi\over \pdr  q_1},p_2={\pdr \phi\over\pdr  q_2}$, we get the equations of state:
\beq
p_1=\left[{2\over 3}\ {q_2-q_2'\over q_1-q_1'}\right]^{3\over 5},\quad p_2=\left[{2\over 3}\ {q_2-q_2'\over q_1-q_1'}\right]^{-{2\over 5}} .\label{eqnstate}
\eeq

 \section{The Wave Equation for a Quantum vdW Gas}
 We can now write down a Schr\"odinger equation by the rule
 \beq
 p_k\to -i{\pd \over \pd q_k}
 \eeq
 familiar from quantum mechanics and optics\footnote{We could easily transform to the more familiar thermodynamic variables through \m{q_1=v, q_2={2\over 3}uv} but then, the equation will look quite complicated. Also, it does not help in solving the equation. So we will continue to work with the `normal co-ordinates'  \m{q_1,q_2}.} 
 \beq
 -i{\pdr^5\over \pdr q_1^2\pdr q_2^3}\psi=\psi
 \eeq

 Recall that the equation we are solving is invariant under translations in $q_1,q_2$. Using this symmetry, we can get the Fourier representation 
  \beq
 \psi(q)=\int e^{{i }\left\{p_1q_1+p_2 q_2\right\}} \delta(p_1^2p_2^3-1)\chi(p_1,p_2) dp_1dp_2  \eeq
with an  arbitrary function $\chi$ on the  curve (analogous to the `mass- shell' of the Klein-Gordon equation)
\beq
p_1^2 p_2^3=1.
\eeq
\begin{figure}
\includegraphics{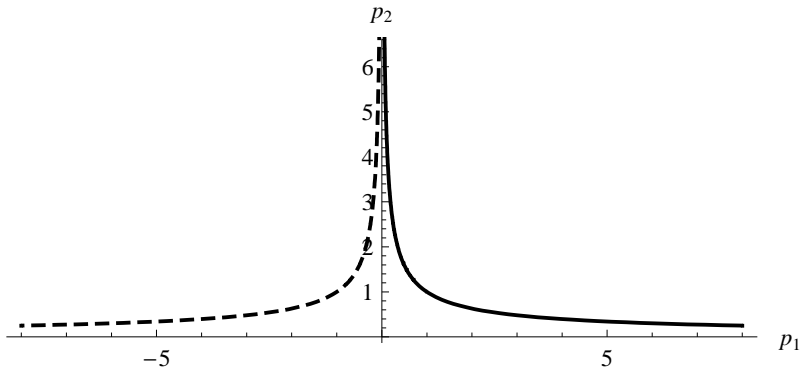}
\caption{The curve $p_1^2 p_2^3=1$. The wavefunctions we study are supported on the branch with $p_1>0$.}
\end{figure}
This curve always has $p_2>0$ and has two branches  with $p_1>0$ and $p_1<0$.  If we look back at the definition of $p_1$ we will see that  positive pressure implies $p_1>0$, which is the branch we will choose. Note also the scaling symmetry
\beq
p_1\to \lambda^3 p_1,\quad p_2\to \lambda^{-2}p_2, \quad\lambda>0.
\eeq
This is analogous to Lorentz transformations that leave the wave equation unchanged. A parametric solution in terms of rational functions of the curve is
\beq
p_1=k^3,\quad p_2=k^{-2}.
\eeq

Any solution of the wave equation supported on the positive branch of the shell  above is of the form 
  
 \beq
 \psi(q_1,q_2)= \int_0^\infty  e^{{i }\left\{q_1 k^3+q_2 k^{-2} \right\}} \tilde\psi(k){dk\over k}.
 \eeq
 We choose ${dk\over k}$ as the measure of integration to respect the scaling symmetry $k\to \lambda k$.
  
 \section{States and Observables}
 
 The quantum Hilbert of the system is the space of solutions to the wave equation. Thus it can be identified with the space of functions $\tilde\psi(k)$ on the shell. It seems reasonable to postulate that the inner product should be
 \beq
 ||\psi||^2=\int_0^\infty |\tilde\psi(k)|^2 {dk\over k}.
 \eeq 
 We have chosen the norm such that the scaling 
 \beq
 \tilde\psi_\lambda(k)=\tilde\psi(\lambda k)
 \eeq
 leaves the norm invariant :
 \beq
 ||\tilde\psi_\lambda||^2=||\tilde\psi||^2.
 \eeq
 
 In this picture, the `pressure' variables (analogous to `momentum' variables of quantum mechanics) are just multiplication operators.
 \beq
 p_1=k^3,\quad p_2=k^{-2}.
 \eeq
 But the `co-ordinate' variables $q_1,q_2$ will be differential operators \cite{NewtonWigner}. Up to additive constants, $q_1=i{\pdr \over \pdr k^3}, q_2=-i{\pdr \over \pdr k^{-2}}$. Taking into account the condition that they be hermitean\footnote{Noice that in this picture, $i\frac{\partial}{\partial k}$ is not hermitean, but $ik\frac{\partial}{\partial k}$ is.} w.r.t.the above inner product, we have\footnote{
  $[A,B]_+=AB+BA$ is the anti-commutator}
 \beq
 \hat q_1=q_1'+{i\over 6}\left[k^{-3}, k{\pdr \over \pdr k}
 \right]_+, \quad \hat q_2=q_2'+{i\over 4}\left[k^{2}, k{\pdr \over \pdr k}
 \right]_+
 \eeq
 The quantities $q_1',q_2'$ are constants  that don't affect the commutators of the operators.

 The expectation value of an observable $\hat A$  represented as a differential operator in $k$ is then
 \beq
 <A>={\int_0^\infty \tilde\psi^*(k){\hat A}\tilde \psi(k){dk\over k}\over
 \int_0^\infty |\tilde\psi(k)|^2 {dk\over k} }.
 \eeq
 
We now look  for a state  that is of finite norm  and is centered at the classical solution. More precisely, the expectation values of the operators above  $p_1,p_2,q_1,q_2$ must satisfy the classical equations of state. 

Note that in this interpretation, the co-ordinate space  wavefunction $\psi(q_1,q_2)$ does not have the meaning of a probability amplitude: in fact it is usually not even square integrable. This is similar to the case with  Klein-Gordon theory \cite{NewtonWigner}.

There could be many such states, but the simplest would be `plane wave' solutions, which are the wave analogues of the fundamental solution of the Hamilton-Jacobi equation. Any other solution of the wave equation would be super-positions of plane waves. For these 
\beq
\tilde\psi(k)=\Theta(k) e^{i [z_1k^3+z_2k^{-2}]}
\eeq
for constants \footnote{The Heaviside function in the expression of $\psi$ is actually not necessary, because by definition of the state space, the wave function is defined only for positive values of $k$.}   $z_1,z_2$. But, as in quantum mechanics, these are not normalizable if $z_1,z_2$ are real. If they have a positive imaginary part, they become square-integrable:
\beq
  {\rm Im}\  z_1,{\rm Im}\  z_2>0.
\eeq
The parameters $z_1,z_2$ can be chosen so that the mean values of observables are centered at classical values: loosely analogous to the coherent states of quantum mechanics \cite{KlauderSudarshan}.

We are thus led to the wave function 
\beq
\tilde\psi(k|\kappa,x,\sigma)= e^{-i x k^3}e^{-{{\kappa^2}\over 10\sigma^2}\left[{1\over 3}\left({k\over \kappa}\right)^3+{1\over 2}\left({{\kappa}\over  k}\right)^2.        \label{wavefn}
\right]}
\eeq
\begin{figure}
\includegraphics{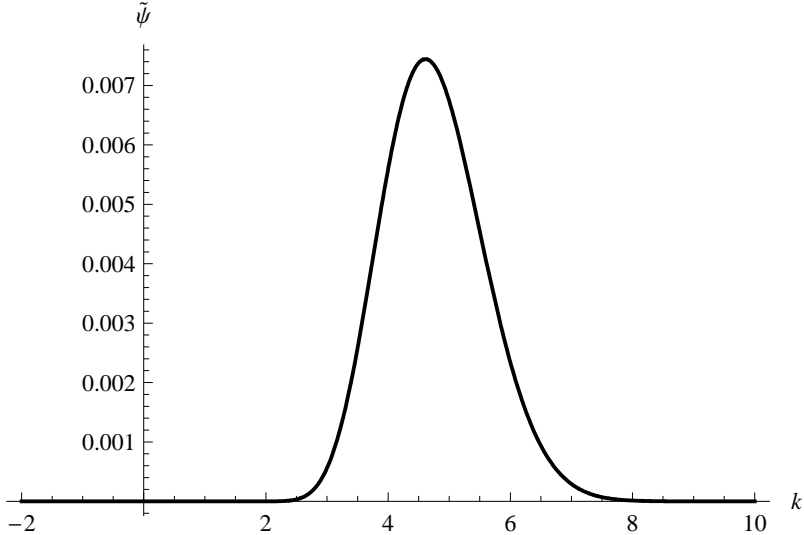}
\caption{The wave function with $x=0,\kappa=5,\sigma=1$}
\end{figure}
\begin{figure}
\includegraphics{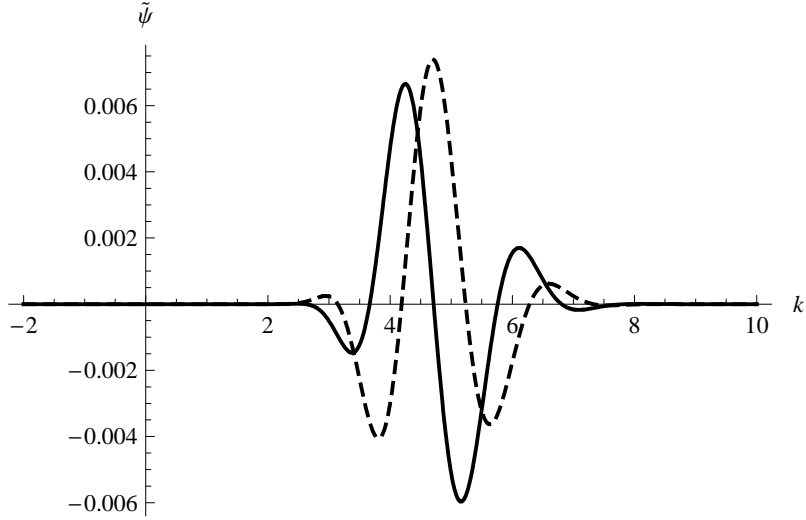}
\caption{The real and imaginary parts of the  wave function for  $x=1,\kappa=5,\sigma=1$}
\end{figure}

This  function is smooth: it vanishes at $k=0$ along with all of its derivatives. The minimum of the quantity in square brackets is at $\kappa$: this is the most probable value of $k$. Near that point, a little calculus gives,
\beq
|\tilde\psi(k)|^2\approx e^{-{(k-\kappa)^2\over 2\sigma^2}}.
\eeq
In the limit of small $\sigma$, we can estimate expectation values using the steepest descent approximation; essentially,  approximate it by the Gaussian:
\beq
<k>\approx \kappa,\quad \Delta k^2=<(k-<k>)^2>\approx \sigma^2.
\eeq

It follows that 
\beq
<p_1>\approx \kappa^3,\quad <p_2>\approx \kappa^{-2}.
\eeq
Also,
\beq
\Delta p_1\approx 3\kappa^2 \sigma,\quad \Delta p_2\approx 2\kappa^{-3}\sigma .
\eeq

In the same approximation,
\beq
<\hat q_1>\approx q_1'+x,\quad <\hat q_2>\approx q_2'+{3\over 2}{\kappa}^5 x.
\eeq
Eliminating $x,{\kappa}$ from the formula for the expectation values, 
\beq
{\kappa}=\left[{2\over 3}\ {<q_2>-q_2'\over <q_1>-q_1'}\right]^{1\over 5}
\eeq
we get the classical equations of state (compare with equation \ref{eqnstate}):
\beq
<p_1>\approx \left[{2\over 3}{<q_2>-q_2'\over <q_1>-q_1'}\right]^{3\over 5}
,\quad
 <p_2>\approx \left[{2\over 3}{<q_2>-q_2'\over <q_1>-q_1'}\right]^{-{2\over 5}}.
\eeq

A little more work shows that 
\beq
\Delta q_1\approx {1\over 6 \kappa^2\sigma},\quad \Delta q_2\approx{\kappa^3\over 4 \sigma}.
\eeq
Thus, for small $\sigma$,
\beq
\Delta p_1 \Delta q_1\approx \half,\quad  \Delta p_2 \Delta q_2\approx \half.
\eeq
Although the probability of small fluctuation is well-approximated by a Gaussian, rare large fluctuations can have a very different probability. If the strategy for observing such fluctuations rely on such rare events, the full wave-function  given above must be used and not just its Gaussian approximation.

Thus we have found a state that is centered at the classical thermodynamic values of the observables, with fluctuations determined by the parameter $\sigma$. We expect that this parameter is a property of the ground state of the molecule that makes up the gas, just like the van der Walls constants $a,b$. For example $b$ is the volume of the molecule: $b^{1\over 3}$ is roughly the mean distance of the outermost electron from the center of mass of the molecule. Similarly ${a\over b}$, which has the dimensions of energy is roughly the position the lowest lying resonance in two body elastic scattering of the molecules.  Just as van der Waals theory itself makes no prediction for $a,b$, we make no prediction within our theory for $\sigma$. However, we should expect that it is roughly the same order of magnitude as $a,b$. For the reasons explained earlier, quantum fluctuations in thermodynamic quantities are suppressed when the thermal wave length is small, so even when $\sigma$ is of the same order as $a,b$, its effect will be measurable only at very small temperatures.

Although $a,b$ used to be thought of as intrinsic properties of molecules, it is possible to manipulate the wavefunction of the outermost electron by external electromagnetic fields. Feshbach  \cite{Feshbach}  resonances can be used to change the scattering lengths by several orders of magnitude, leading to the same change in $a$. Perhaps such clever experimental techniques will be found some day to enhance the value of $\sigma$ as well.

It would be most interesting to measure experimentally the quantum fluctuations in thermodynamic quantities. In the theoretical direction, the analogy with Klein-Gordon theory suggests a second-quantization of our wave equation. We do not know  yet what that would mean physically.
  
\section{Acknowledgement}
We thank Anosh Joseph for discussions.  This work was supported in part by the Department of Energy under the contract number DE-FG02-91ER40685.

\section{Appendix: The van der Waals Constants}

Tables of van der Waals constants \cite{vdWTable} are easily available.
We  recall a few values to give an idea of the order of magnitudes.

For hydrogen molecules, $b=0.027\times 10^{-3} {{\rm m}^3\over {\rm mol}}$. Using Avogadro's number, $N_A=6\times 10^{23}$ we get 
\beq
b={0.027\over 6}\times 10^{-26} {\rm m}^3\approx\left(3.56\times10^{-10} {\rm m} \right)^3
\eeq
Thus $b$ is of the order of the volume of a cube of side equal to $7.1$ Bohr radii.

Also, for hydrogen molecules, $a=0.025 {{\rm J}{\rm m}^3\over {\rm mol}^2}$. Thus ${a\over b}\approx 1{{\rm J}\over {\rm mol}}\approx 10 {\mu}{\rm eV}$ which is much smaller than atomic energy levels.

For Helium, $b=0.024\times10^{-3} {{\rm m}^3\over {\rm mol}}$, so that $b^{1\over 3}\approx 6.8 a_B$ and $a=0.0034 {{\rm J}{\rm m}^3\over {\rm mol}^2}$, so that ${a\over b}\approx 1.5 {\mu} {eV}.$ As expected, Helium atoms interact much more weakly than Hydrogen molecules.

\pagebreak

\end{document}